\documentclass[aps,prb,twocolumn,showpacs,floatfix]{revtex4}

\usepackage{graphicx}

\begin{document}

\title{Quasiparticle states of the Hubbard model near the Fermi level}

\author{A.~Sherman}

\affiliation{Institute of Physics, University of Tartu, Riia 142, 51014
Tartu, Estonia}

\date{\today}

\begin{abstract}
The spectra of the $t$-$U$ and $t$-$t'$-$U$ Hubbard models are
investigated in the one-loop approximation for different values of the
electron filling. It is shown that the four-band structure which is
inherent in the case of half-filling and low temperatures persists also
for some excess or deficiency of electrons. Besides, with some
departure from half-filling an additional narrow band of quasiparticle
states arises near the Fermi level. The dispersion of the band, its
bandwidth and the variation with filling are close to those of the
spin-polaron band of the $t$-$J$ model. For moderate doping spectral
intensities in the new band and in one of the inner bands of the
four-band structure decrease as the Fermi level is approached which
leads to the appearance of a pseudogap in the spectrum.
\end{abstract}

\pacs{71.10.Fd, 71.27.+a}

\maketitle

\section{Introduction}
In the last two decades the discovery of high-$T_c$ superconductors,
heavy-fermion compounds and organic conductors has revived interest in
strongly correlated electron systems. One of the simplest and still
realistic models in this field is the one-band Hubbard model
two-dimensional version of which has been extensively studied in
connection with the cuprate perovskite superconductors. Along with
Monte-Carlo simulations, \cite{Moreo,Grober} different cluster methods,
\cite{Maier,Aichhorn,Tremblay} the operator projection technique,
\cite{Mancini} the generating functional approach, \cite{Izyumov05}
various versions of the diagram technique
\cite{Zaitsev,Izyumov,Ovchinnikov,Vladimir,Metzner,Pairault} are used
for the investigation of the model. In a system in which the Coulomb
interaction dominates, it is reasonable to treat this interaction
exactly and the kinetic energy in the framework of a perturbation
theory.

In the diagram technique proposed in
Refs.~\onlinecite{Vladimir,Metzner,Pairault} the power expansion is
expressed in terms of site cumulants of electron creation and
annihilation operators. From the expansion the Larkin equation for the
electron Green's function can be derived.
\cite{Vladimir,Pairault,Sherman06} The equation can be solved in the
one-loop approximation. However, the obtained solution has a flaw -- a
negative spectral weight in two narrow frequency regions.
\cite{Pairault} This flaw can be remedied by an interpolation using
results for regular regions. \cite{Sherman06} The obtained spectral
function \cite{Sherman06} was shown to be in agreement with results of
Monte-Carlo simulations \cite{Grober} at half-filling and for moderate
temperatures when the magnetic correlation length is comparable with
the intersite distance. In particular, it was shown that in agreement
with results of Monte-Carlo and cluster methods
\cite{Moreo,Grober,Maier,Aichhorn,Tremblay} the diagram technique is
able to describe the four-band structure of the spectrum at
half-filling.

In the present paper the same method is used for the investigation of
the energy spectrum at a departure from half-filling. It is shown that
in this case the above-mentioned four bands persist and additionally a
new band arises in some vicinity of the Fermi level. By its properties
-- the dispersion, bandwidth and the variation with filling -- the band
resembles the spin-polaron band of the $t$-$J$ model. \cite{Dagotto}
For moderate doping the spectral intensities in the new band and in one
of the inner bands of the four-band structure decrease as the Fermi
level is approached. This produces a pseudogap near the Fermi level.
For the hole-doped case, $\bar{n}<1$, the magnitude of the pseudogap
observed in photoemission decreases with increasing the hole doping
$1-\bar{n}$, while for the electron-doped case, $\bar{n}>1$, this
magnitude increases with increasing the electron doping $\bar{n}-1$.
Here $\bar{n}$ is the electron concentration. Together with the $t$-$U$
Hubbard model the $t$-$t'$-$U$ model is also considered for the ratio
$t'/t=-0.3$ of the next-nearest and nearest neighbor hopping constants.
\cite{Korshunov} As for the case of half-filling the calculated
spectral functions and dispersions appear to be similar to those
obtained by Monte-Carlo and cluster methods, provided that doping or
temperature are high enough to ensure a short magnetic correlation
length.

Main formulas used in the calculations are given in the following
section (the detailed derivation of these formulas can be found in
Ref.~\onlinecite{Sherman06}). The discussion of the obtained results
and their comparison with results of other methods are carried out in
Sec.~III and~IV for the $t$-$U$ and $t$-$t'$-$U$ models, respectively.
Concluding remarks are presented in Sec.~V\@.

\section{Main formulas}
The Hamiltonian of the Hubbard model reads
\begin{equation}\label{Hamiltonian}
H=\sum_{\bf nm\sigma}t_{\bf nm}a^\dagger_{\bf n\sigma}a_{\bf m\sigma} +
\frac{U}{2}\sum_{\bf n\sigma}n_{\bf n\sigma}n_{\bf n,-\sigma},
\end{equation}
where $t_{\bf nm}$ is the hopping constants, the operator
$a^\dagger_{\bf n\sigma}$ creates an electron on the site {\bf n} of
the two-dimensional square lattice with the spin projection $\sigma=\pm
1$, $U$ is the on-site Coulomb repulsion and the electron number
operator $n_{\bf n\sigma}=a^\dagger_{\bf n\sigma}a_{\bf n\sigma}$.

\begin{figure}[t]
\includegraphics[width=8.2cm]{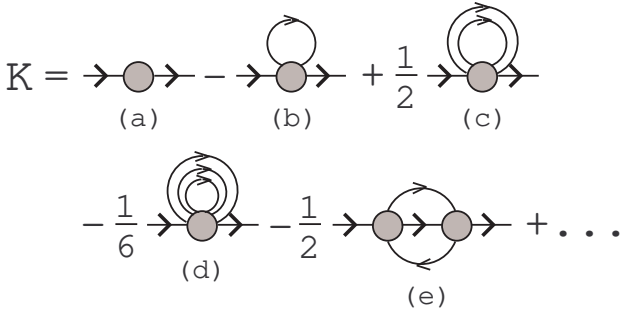}
\caption{Irreducible diagrams of the first four orders of the expansion
in powers of $t_{\bf nm}$.}\label{Fig_i}
\end{figure}

The diagram technique proposed in Refs.~\onlinecite{Vladimir} and
\onlinecite{Metzner} is used in the present work for calculating
Green's function
\begin{equation}\label{GF}
G({\bf n'\tau',n\tau})=\langle{\cal T}\bar{a}_{\bf n'\sigma}(\tau')
a_{\bf n\sigma}(\tau)\rangle,
\end{equation}
where the angular brackets denote the statistical averaging with the
Hamiltonian ${\cal H}=H-\mu\sum_{\bf n\sigma}n_{\bf n\sigma}$, $\mu$ is
the chemical potential, ${\cal T}$ is the time-ordering operator which
arranges other operators from right to left in ascending order of times
$\tau$, $a_{\bf n\sigma}(\tau)=\exp({\cal H}\tau)a_{\bf
n\sigma}\exp(-{\cal H}\tau)$, and $\bar{a}_{\bf
n\sigma}(\tau)=\exp({\cal H}\tau)a^\dagger_{\bf n\sigma}\exp(-{\cal
H}\tau)$. With the use of the diagram technique the following Larkin
equation is derived \cite{Vladimir,Pairault,Sherman06} for the Fourier
transform of function~(\ref{GF}):
\begin{equation}\label{Larkin}
G({\bf k},i\omega_l)=\frac{K({\bf k},i\omega_l)}{1-t_{\bf k}K({\bf
 k},i\omega_l)},
\end{equation}
where $t_{\bf k}=\sum_{\bf n}e^{-i{\bf k(n-m)}}t_{\bf nm}$,
$\omega_l=(2l+1)\pi T$ is the Matsubara frequency with the temperature
$T$ and an integer $l$, $K({\bf k},i\omega_l)$ is the sum of all
irreducible diagrams -- the diagrams which cannot be divided into two
parts by cutting a hopping line. Such diagrams which appear in the
first four orders of the perturbation expansion in powers of $t_{\bf
nm}$ are shown in Fig.~\ref{Fig_i} with their signs and prefactors.
Here circles are cumulants \cite{Kubo} of electron operators. In
diagrams, the order of a cumulant is determined by a number of incoming
or outgoing hopping lines (directed lines in Fig.~\ref{Fig_i}).
Cumulants of the first and second orders which will be used below read
\begin{eqnarray*}
&&K_1(\tau'\sigma',\tau\sigma)=\langle{\cal T}\bar{a}_\sigma(\tau')
 a_\sigma(\tau)\rangle_0\delta_{\sigma\sigma'},\nonumber\\
&&K_2(\tau'\sigma,\tau\sigma,\tau'_1\sigma_1,\tau_1\sigma_1)=\nonumber\\
&&\quad\langle{\cal T}\bar{a}_\sigma(\tau') a_\sigma(\tau)
 \bar{a}_{\sigma_1}(\tau'_1) a_{\sigma_1}(\tau_1)\rangle_0\\
&&\quad -K_1(\tau'\sigma,\tau\sigma)K_1(\tau'_1\sigma_1,\tau_1\sigma_1)
 \nonumber\\
&&\quad +K_1(\tau'\sigma,\tau_1\sigma_1)K_1(\tau'_1\sigma_1,\tau
 \sigma), \nonumber
\end{eqnarray*}
where the subscript ``0'' near the angular bracket indicates that the
averaging and time dependencies of the operators are determined by the
site Hamiltonian $H_{\bf n}=\sum_\sigma[ (U/2)n_{\bf n\sigma}n_{\bf
n,-\sigma}-\mu n_{\bf n\sigma}]$. All operators in the cumulants belong
to the same lattice site. Due to the translational symmetry the
cumulants do not depend on the site index which is therefore omitted in
the above equations.

Partial summation is implied in the diagrams in Fig.~\ref{Fig_i} -- the
irreducible diagrams are included in the hopping lines which therefore
correspond to the expression
\begin{equation}\label{hopping}
\Theta({\bf k},i\omega_l)=\frac{t_{\bf k}}{1-t_{\bf k}K({\bf
 k},i\omega_l)}=t_{\bf k}+t_{\bf k}^2G({\bf k},i\omega_l).
\end{equation}

In the one-loop approximation used below, the total collection of
irreducible diagrams $K({\bf k},i\omega_l)$ is substituted by the sum
of the two diagrams (a) and (b) in Fig.~\ref{Fig_i}. Thus,
\begin{eqnarray}
K(i\omega_l)&=&K_1(i\omega_l)
  -T\sum_{l_1\sigma_1}K_2(i\omega_l\sigma,i\omega_{l_1}\sigma_1,
  i\omega_{l_1}\sigma_1)\nonumber\\
&&\times\frac{1}{N}\sum_{\bf k}t^2_{\bf k}G({\bf
  k},i\omega_{l_1}),\label{oneloop}
\end{eqnarray}
where $K_1(i\omega_l)$ and $K_2(i\omega_l\sigma,i\omega_{l_1}\sigma_1,
i\omega_{l_1}\sigma_1)$ are the Fourier transforms of the cumulants of
the first and second orders, respectively, $N$ is the number of sites
and I set $\sum_{\bf k}t_{\bf k}=0$. Notice that in this approximation
$K$ does not depend on momentum. The cumulants read \cite{Sherman06}
\begin{eqnarray}
&&K_1(i\omega_l)=\frac{1}{Z_0}\left(\frac{e^{-\beta E_1}+e^{-\beta
 E_0}}{i\omega_l-E_{10}}+\frac{e^{-\beta E_2}+e^{-\beta
 E_1}}{i\omega_l-E_{21}}\right),\nonumber\\
&&\sum_{\sigma_1}K_2(i\omega_l\sigma,i\omega_{l_1}\sigma_1,
 \omega_{l_1}\sigma_1)=-Z_0^{-1}U \bigl\{e^{-\beta E_0}g_{01}
 (i\omega_l)\nonumber \\
&&\quad\times g_{01}(i\omega_{l_1})g_{02}(i\omega_l+
 i\omega_{l_1})\bigl[g_{01}(i\omega_l)+g_{01}
 (i\omega_{l_1})\bigr] \nonumber\\
&&\quad+e^{-\beta E_2}g_{12}(i\omega_l)g_{12}
 (i\omega_{l_1})g_{02}(i\omega_l+i\omega_{l_1})\bigl[
 g_{12}(i\omega_l)\nonumber\\
&&\quad+g_{12}(i\omega_{l_1})
 \bigr]+e^{-\beta E_1}\bigl[g_{01}(i\omega_l)g_{12}
 (i\omega_l)\bigl(g_{01}(i\omega_{l_1}) \nonumber\\[-1ex]
&&\label{cumulants}\\[-1ex]
&&\quad-g_{12}(i\omega_{l_1})\bigr)^2+
 g_{01}(i\omega_{l_1})g_{12}(i\omega_{l_1})
 \bigl(g_{01}^2(i\omega_l)\nonumber\\
&&\quad+g_{12}^2(i\omega_l)\bigr)\bigr]\bigr\}-Z_0^{-2}
 U^2\beta\delta_{ll_1}\bigl(e^{-\beta(E_0+E_2)} \nonumber\\
&&\quad+2e^{-\beta(E_0+E_1)}+3e^{-2\beta
 E_1}+2e^{-\beta(E_1+E_2)}\bigr)
 g_{01}^2(i\omega_l)\nonumber\\
&&\quad\times g_{12}^2(i\omega_l)+Z_0^{-2}U^2\beta
 \bigl(2e^{-\beta(E_0+E_2)}+e^{-\beta(E_0+E_1)} \nonumber\\
&&\quad+e^{-\beta(E_1+E_2)}\bigr)g_{01}(i\omega_l)g_{12}
 (i\omega_l)g_{01}(i\omega_{l_1})g_{12}(i\omega_{l_1}),
 \nonumber
\end{eqnarray}
where $\beta=T^{-1}$, $E_0=0$, $E_1=-\mu$, and $E_2=U-2\mu$ are the
eigenenergies of the site Hamiltonian $H_{\bf n}$, $E_{ij}=E_i-E_j$,
$Z_0=e^{-\beta E_0}+2e^{-\beta E_1}+e^{-\beta E_2}$ is the site
partition function, $g_{ij}(i\omega_l)=(i\omega_l+ E_{ij})^{-1}$.

Equations~(\ref{cumulants}) can be significantly simplified for the
case of principal interest $U\gg T$. In this case if $\mu$ satisfies
the condition
\begin{equation}\label{condition}
\lambda<\mu<U-\lambda,
\end{equation}
where $\lambda\gg T$, the exponent $e^{-\beta E_1}$ is much larger than
$e^{-\beta E_0}$ and $e^{-\beta E_2}$. By passing to real frequencies
one can ascertain that terms in $\sum_{\sigma_1}K_2$ with the two
latter multipliers contain the same peculiarities as terms with
$e^{-\beta E_1}$. Therefore terms with $e^{-\beta E_0}$ and $e^{-\beta
E_2}$ can be omitted and Eq.~(\ref{cumulants}) is simplified to
\begin{eqnarray}
&&K_1(i\omega_l)=\frac{i\omega_l+\mu-U/2}{(
 i\omega_l+\mu)(i\omega_l+\mu-U)},\label{simplified_i}\\
&&\sum_{\sigma_1}K_2(i\omega_l\sigma,i\omega_{l_1}\sigma_1,
 i\omega_{l_1}\sigma_1)\nonumber\\
&&\quad=-\frac{1}{2}Ug_{01}(i\omega_l)g_{12}(
 i\omega_l) \Bigl[g_{01}^2(i\omega_{l_1})+g_{1
 2}^2(i\omega_{l_1})\Bigr]\nonumber\\
&&\quad\quad-\frac{1}{2}Ug_{01}(i\omega_{l_1})g_{1
 2}(i\omega_{l_1})\Bigl[g_{01}(i\omega_l)-
 g_{12}(i\omega_l)\Bigr]^2\nonumber\\
&&\quad\quad-\frac{3}{4}U^2\beta
 \delta_{ll_{1}}g_{01}^2(i\omega_l)
 g_{12}^2(i\omega_l).\label{simplified_ii}
\end{eqnarray}

Further simplification can be achieved by using the Hubbard-I
approximation \cite{Hubbard} for the Green's function $G({\bf
k},i\omega_{l_1})$ on the right-hand side of Eq.~(\ref{oneloop}). The
respective expression is derived from Eq.~(\ref{Larkin}) if the total
irreducible part $K({\bf k},i\omega_l)$ is approximated by the first
cumulant $K_1(i\omega_l)$ [the diagram (a) in Fig.~\ref{Fig_i}] from
Eqs.~(\ref{cumulants}) or~(\ref{simplified_i}).
\cite{Izyumov,Vladimir,Pairault} This gives
\begin{eqnarray}
G({\bf k},i\omega_l)&=&\frac{1}{2}\left(1+\frac{t_{\bf
 k}}{\sqrt{U^2+t_{\bf k}^2}}\right)\frac{1}{i\omega_l-
 \varepsilon_{1,\bf k}}\nonumber\\
 &&+\frac{1}{2}\left( 1-\frac{t_{\bf k}}{\sqrt{U^2+t_{\bf k}^2}}
 \right)\frac{1}{i\omega_l-\varepsilon_{2,\bf k}},\nonumber\\[-.5ex]
&&\label{HubbardI}\\[-.5ex]
\varepsilon_{1,\bf k}&=&\frac{1}{2}\left(U+t_{\bf k}+\sqrt{U^2+t_{\bf
 k}^2}\right)-\mu,\nonumber\\
\varepsilon_{2,\bf k}&=&\frac{1}{2}\left(U+t_{\bf k}-
 \sqrt{U^2+t_{\bf k}^2}\right)-\mu.\nonumber
\end{eqnarray}

After carrying out the summation over $l_1$ in Eq.~(\ref{oneloop}) with
the use of Eqs.~(\ref{simplified_i})--(\ref{HubbardI}) it is convenient
to turn to real frequencies by substituting $i\omega_l$ with
$z=\omega+i\eta$ where $\eta$ is a small positive constant which
affords an artificial broadening.

\section{The $t$-$U$ model}
At first let us consider the $t$-$U$ model in which only the nearest
neighbor hopping constant $t$ is nonzero and $t_{\bf
k}=2t[\cos(k_x)+\cos(k_y)]$. Here the intersite distance is taken as
the unit of length. Due to the electron-hole symmetry in this case the
consideration can be restricted to the range of the chemical potentials
$\mu\leq U/2$.

\begin{figure}
\includegraphics[width=7cm]{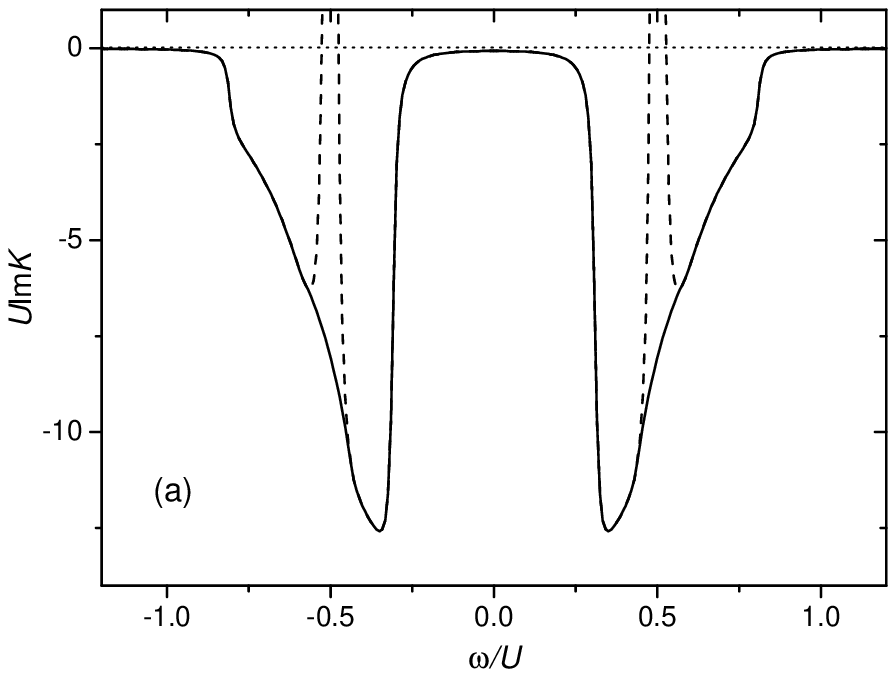}

\vspace{4ex}
\includegraphics[width=7.3cm]{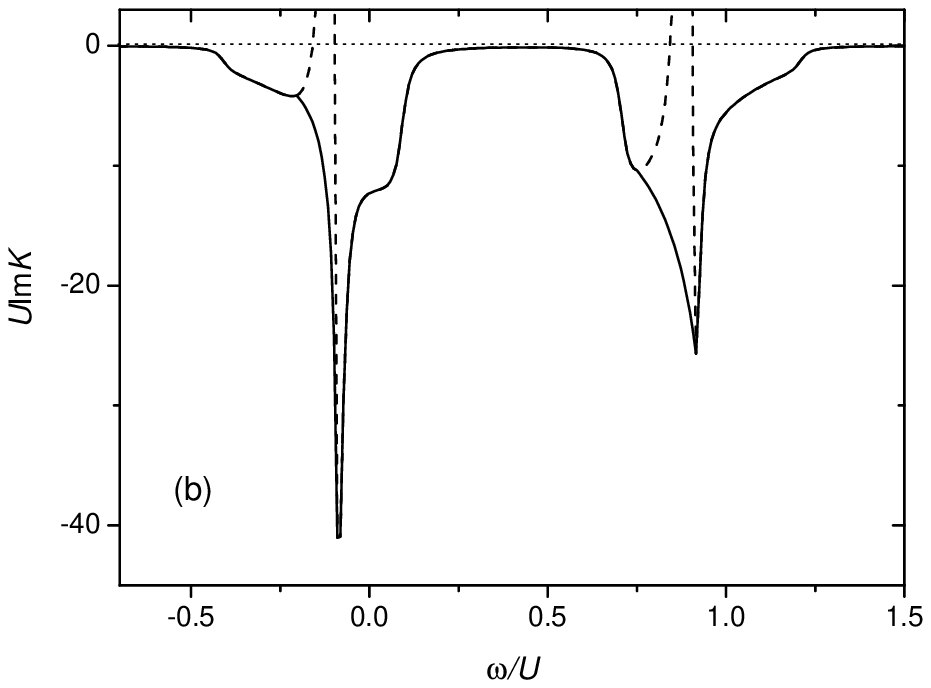} \caption{The imaginary part of
$K(\omega)$ calculated using Eqs.~(\protect\ref{oneloop}) and
(\protect\ref{simplified_i})--(\protect\ref{HubbardI}) for a
100$\times$100 lattice, $t=-U/8$ and $T=0.001U$ (the dashed lines). (a)
$\mu=0.5U$, $\eta=0.01U$. (b) $\mu=0.1U$, $\eta=0.02U$. The solid lines
show the corrected ${\rm Im} K(\omega)$ (see text).} \label{Fig_ii}
\end{figure}
Figure~\ref{Fig_ii} demonstrates ${\rm Im} K(\omega)$ calculated with
the use of Eqs.~(\ref{oneloop}) and
(\ref{simplified_i})--(\ref{HubbardI}). The change to real frequencies
converts the Matsubara function~(\ref{GF}) into the retarded Green's
function. \cite{Abrikosov} It is an analytic function in the upper
half-plane which requires that ${\rm Im} K(\omega)$ be negative. As
seen from Fig.~\ref{Fig_ii}, this condition is violated at
$\omega_d=-\mu$ and $U-\mu$. The problem is connected with divergencies
at these frequencies introduced by functions $g_{01}(\omega)$ and
$g_{12}(\omega)$ in the above formulas. As can be seen from the
procedure of calculating the cumulants, these functions and
divergencies with sign-changing residues will appear in all orders of
the perturbation theory. It can be expected that in the entire series
the divergencies compensate each other and the resulting ${\rm Im}
K(\omega)$ is negative everywhere. However, in the considered subset of
terms such compensation does not occur. Nevertheless, as seen from
Fig.~\ref{Fig_ii}, at frequencies neighboring to $\omega_d$ the
irreducible part is regular and, if the used subset of diagrams is
expected to give a correct estimate of the entire series for these
frequencies, the values of ${\rm Im} K(\omega)$ near $\omega_d$ can be
reconstructed using an interpolation and its values in the regular
region. \cite{Sherman06} Examples of such interpolation are given in
Fig.~\ref{Fig_ii}. The function $K(z)$ has to be analytic in the upper
half-plane also and therefore its real part can be calculated from its
imaginary part using the Kramers-Kronig relations. I used this way with
the interpolated ${\rm Im} K(\omega)$ to avoid the influence of the
divergencies on ${\rm Re} K(\omega)$. However, the application of the
interpolation overrates somewhat values of $|{\rm Im} K(\omega)|$ which
leads to the overestimation of the tails in the real part. To correct
this defect the interpolated $K(\omega)$ is scaled so that in the far
tails its real part coincides with the values obtained from
equation~(\ref{oneloop}).

\begin{figure*}
\includegraphics[width=7cm]{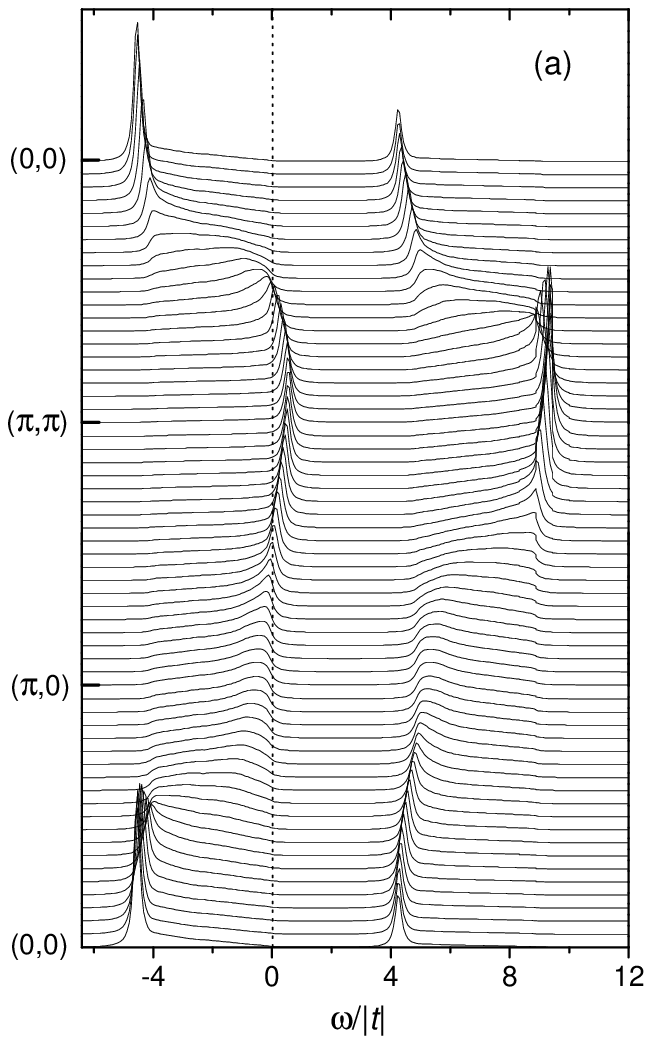}\hspace{2em}
\raisebox{-.55ex}{\includegraphics[width=7.15cm]{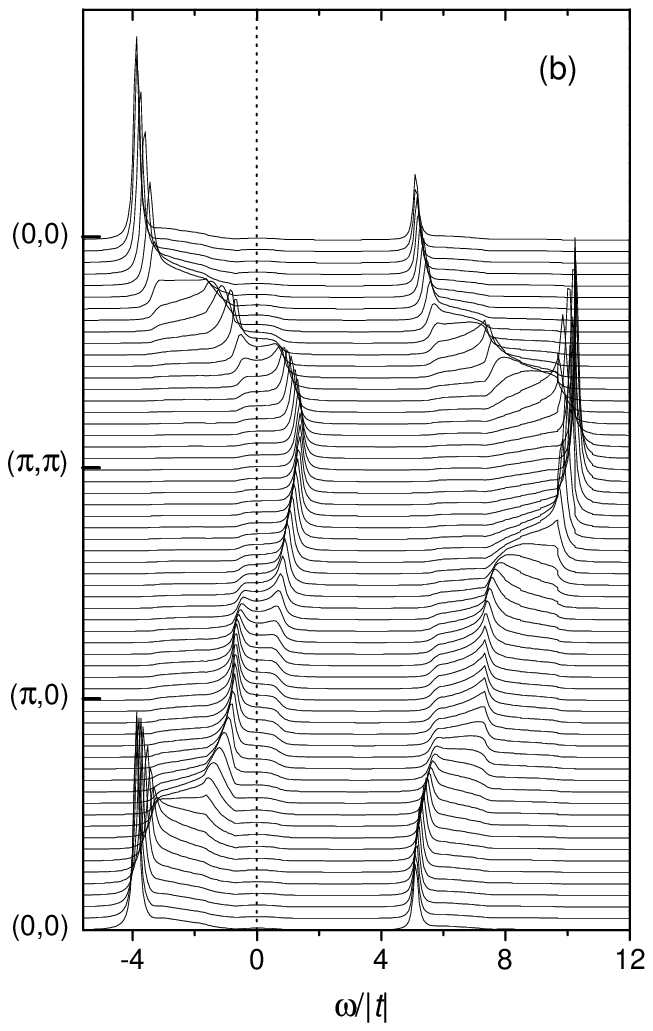}}
\caption{The spectral function $A({\bf k}\omega)$ of the $t$-$U$ model
calculated for momenta along the symmetry lines of the square Brillouin
zone in a 40$\times$40 lattice for $t=-U/8$, $T=0.001U$, $\eta=0.02U$,
$\mu=0.2U$ (a), and $\mu=0.1U$ (b).} \label{Fig_iii}
\end{figure*}
As seen from Fig.~\ref{Fig_ii}a, at half-filling, $\mu=U/2$, ${\rm Im}
K(\omega)$ has two broad minima. With decreasing the chemical potential
from this value the minima shift with respect to the Fermi level
without a noticeable change of their shapes until the Fermi level
enters the left minimum which for $t=-U/8$ occurs at $\mu\approx
0.17U$. As this takes place, two new sharp minima arise near
frequencies $-\mu$ and $U-\mu$ on the background of the above-mentioned
broad minima (see Fig.~\ref{Fig_ii}b). The appearance of the broad
features in Fig.~\ref{Fig_ii} is connected with the third term on the
right-hand side of Eq.~(\ref{simplified_ii}), while the sharp minima
are related to the second term in this formula. Its contribution to
$K(\omega)$, Eq.~(\ref{oneloop}), grows rapidly when the Fermi level
enters the broad minimum.

The spectral function
\begin{eqnarray}
A({\bf k}\omega)&=&-\frac{1}{\pi}{\rm Im}G({\bf k}\omega)\nonumber\\
&=&-\frac{1}{\pi}\frac{{\rm Im}K(\omega)}{[1-t_{\bf k}{\rm Re}
 K(\omega)]^2+[t_{\bf k}{\rm Im}K(\omega)]^2}\label{specfun}
\end{eqnarray}
obtained from such calculated irreducible part for momenta along the
symmetry lines of the square Brillouin zone is shown in
Fig.~\ref{Fig_iii}. The shapes of the spectral function in
Fig.~\ref{Fig_iii}a are nearly the same as at half-filling
\cite{Sherman06} -- as in the case of $K(\omega)$, with decreasing
$\mu$ from $U/2$ to $0.17U$ these curves shift with respect to the
Fermi level without perceptible changes in their shape. In agreement
with results of Monte-Carlo simulations \cite{Moreo,Grober} and cluster
methods \cite{Maier,Aichhorn,Tremblay} four bands can be distinguished
in these spectra. For parameters of Fig.~\ref{Fig_iii}a these bands are
located near the frequencies $-4|t|$, 0, $4|t|$, and $9|t|$. For the
major part of the Brillouin zone maxima forming the bands arise at
frequencies which satisfy the equation $1-t_{\bf k}{\rm Re}K(\omega)=0$
and fall into the region of a small damping $|{\rm Im} K(\omega)|$ [see
Eq.~(\ref{specfun})]. As seen from Fig.~\ref{Fig_ii}, such regions of
small damping are located between and on the outside of the two broad
minima in ${\rm Im}K(\omega)$. This is the reason of the existence of
the four well separated bands -- two of them are located between the
minima of ${\rm Im}K(\omega)$, while two others are on the outside of
these minima. Broader maxima of $A({\bf k}\omega)$ for momenta near the
boundary of the magnetic Brillouin zone are of different nature --
since $t_{\bf k}$ is small for such momenta, the resonant denominator
in Eq.~(\ref{specfun}) does not vanish and the broad maxima in the
spectral functions reproduce the maxima of $-{\rm Im}K(\omega)$ in the
numerator of this formula.

More substantial changes in $A({\bf k}\omega)$ occur for $\mu\alt
0.17U$. As seen from Fig.~\ref{Fig_iii}b, the four-band structure
persists also in this case. In addition to this there appear sharp
dispersive features near $\omega=-\mu$ and $U-\mu$, the latter feature
being substantially weaker in the case of hole doping, $\mu<U/2$. It is
clear that these changes in the spectral function are connected with
the changes in ${\rm Im}K(\omega)$ shown in Fig.~\ref{Fig_ii}. For the
hole-doping case the peaks near $-\mu$ are located in the nearest
vicinity of the Fermi level. For the parameters of Fig.~\ref{Fig_iii}b
neither these peaks nor the peaks forming the lower inner band cross
the Fermi level -- their intensities decrease as it is approached. As a
consequence a pseudogap arises in the spectrum near the Fermi level.
Recently spectra with analogous pseudogaps were also obtained by
cluster methods. \cite{Maier,Tremblay,Senechal,Kyung,Macridin} In these
works such spectral peculiarities were identified with the pseudogap
observed in the photoemission of cuprates. \cite{Damascelli}

\begin{figure}
\includegraphics[width=7cm]{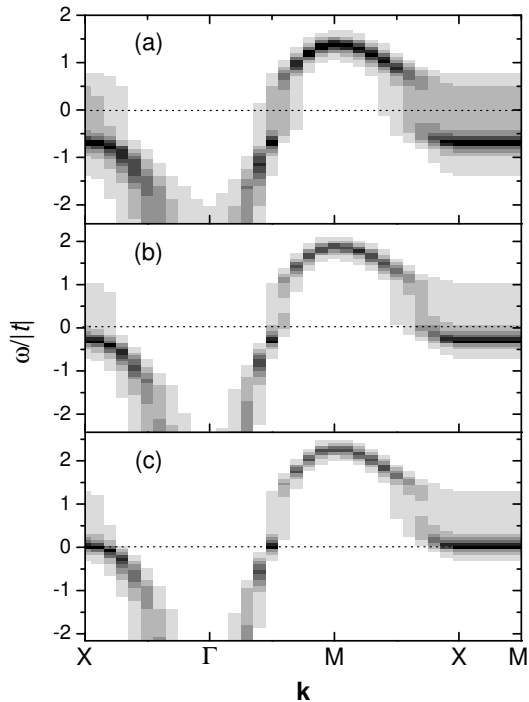}
\caption{The dispersion of maxima of the spectral function in the
$t$-$U$ model on a 20$\times$20 lattice for $t=-U/8$, $T=0.001U$,
$\eta=0.02U$, $\mu=0.1U$ (a), $0.05U$ (b), and $0.01U$ (c). Here darker
areas correspond to larger intensities. The points X, $\Gamma$, M, and
M' correspond to the momenta $(\pi,0)$, $(0,0)$, $(\pi,\pi)$, and
$(\pi/2,\pi/2)$, respectively.} \label{Fig_iv}
\end{figure}
Figure~\ref{Fig_iv} demonstrates the dispersion of maxima of the
spectral function near the Fermi level for three values of the chemical
potential. These maxima form the lover inner band above the Fermi level
and the new band arising at $\mu\approx 0.17U$ below and at the Fermi
level. At this value of the chemical potential the width of the new
band is approximately equal to $|t|=2J$ where $J=4t^2/U$ is the
superexchange constant of the effective Heisenberg model which
describes magnetic excitations in the limit $U\gg|t|$.
\cite{Izyumov,Dagotto} The bandwidth decreases with reduction in the
electron concentration. The maximum energies of the band are located
near the boundary of the magnetic Brillouin zone. The dispersion is
much larger in the direction $(\pi/2,\pi/2)-(0,0)$ than along the
boundary of the magnetic Brillouin zone $(\pi,0)-(0,\pi)$. These
properties of the band resembles those of the spin-polaron band of the
$t$-$J$ model. This latter band is also located near the Fermi level,
has the similar dispersion and the bandwidth, which decreases with
decreasing $\bar{n}$. \cite{Dagotto,Sherman94} As seen from
Fig.~\ref{Fig_iv}, with decreasing $\mu$ and $\bar{n}$ the Fermi level
shifts to the lower edge of the pseudogap and enters the new band at
$\mu\approx 0.07U$. In the photoemission which probes the part of the
spectral function occupied by electrons this change in the energy
spectrum will look like the decrease with the subsequent disappearance
of the pseudogap with decreasing the electron concentration. Such
behavior of the pseudogap is indeed observed in hole-doped cuprates.
\cite{Damascelli} Notice that in the case of electron doping the
pseudogap observed in photoemission will grow with $\bar{n}$ until the
Fermi level enters the new band (see Fig.~\ref{Fig_vii}).

\begin{figure}
\includegraphics[width=6cm]{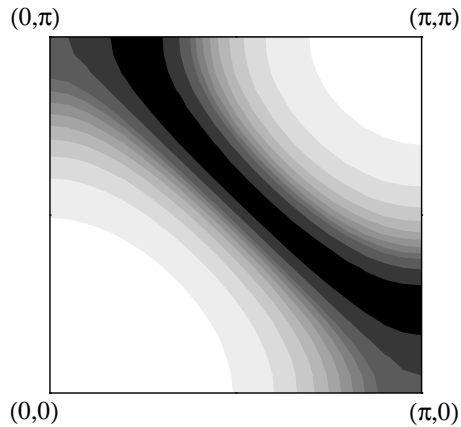}
\caption{The intensity plot of the spectral function at the Fermi level
in the first quadrant of the Brillouin zone. The darker areas
correspond to larger intensities. The $t$-$U$ model on a 40$\times$40
lattice with the parameters $t=-U/8$, $T=0.001U$, $\eta=0.02U$, and
$\mu=0.05U$.} \label{Fig_v}
\end{figure}

The intensity plot of the spectral function at the Fermi level is shown
in Fig.~\ref{Fig_v}. The plot was obtained by averaging the spectral
function in the frequency range $[-0.01U,+0.01U]$. The dark area which
corresponds to the maximum intensity can be interpreted as the Fermi
surface in the region of the chemical potential $0.12U\alt\mu\alt
0.25U$ where the lower inner band crosses the Fermi level (see
Fig.~\ref{Fig_iii}a), and for $\mu\alt 0.07U$ where the crossing occurs
with the new band (see Figs.~\ref{Fig_iv}b and~c). For $\mu$ in the
intermediate region maxima of both bands lose their intensities as the
Fermi level is approached (see Figs.~\ref{Fig_iii}b and~\ref{Fig_iv}a).
However, also in this case the intensity plot is similar to that shown
in Fig.~\ref{Fig_v}.
With decreasing $\mu$ and $\bar{n}$ the Fermi surface shrinks to the
center of the Brillouin zone and near $\mu=0.01U$ changes its shape
from a diamond centered at $(\pi,\pi)$ to that centered at $(0,0)$. In
contrast to the results of the cluster methods
\cite{Senechal,Kyung,Macridin} in the used approximation the variation
of the intensity along the Fermi surface is small both in the $t$-$U$
and $t$-$t'$-$U$ models.

\begin{figure*}
\includegraphics[width=7.2cm]{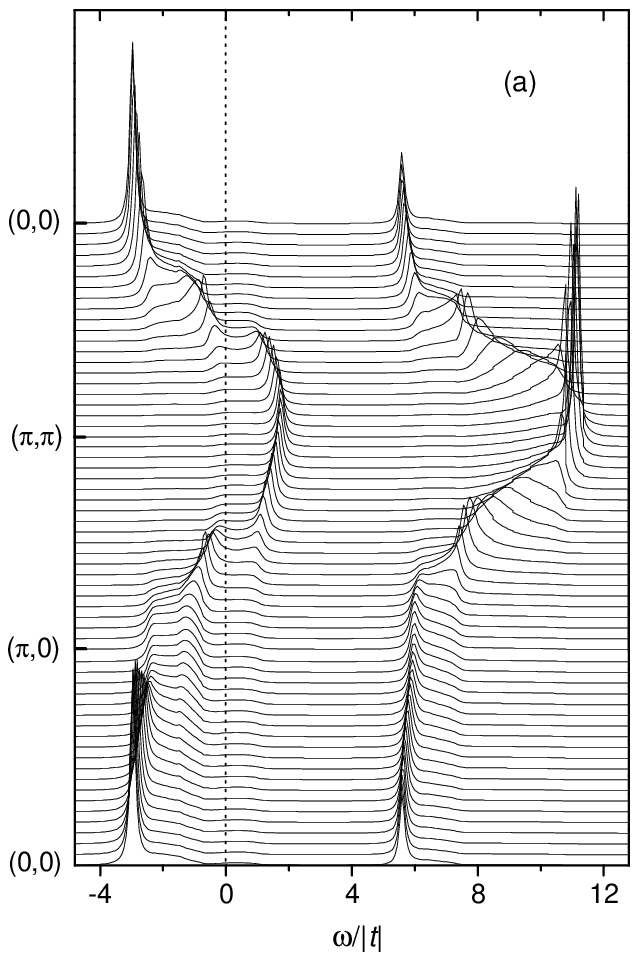}\hspace{2em}
\raisebox{.8ex}{\includegraphics[width=6.95cm]{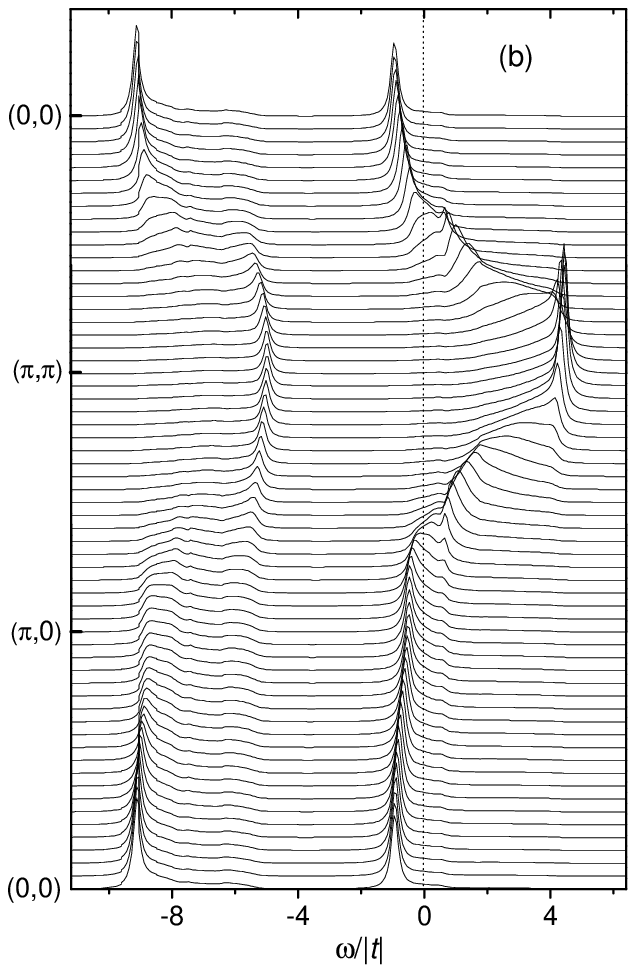}} \caption{The
spectral function $A({\bf k}\omega)$ of the $t$-$t'$-$U$ model
calculated for momenta along the symmetry lines of the square Brillouin
zone in a 40$\times$40 lattice for $t=-U/8$, $t'/t=-0.3$, $T=0.001U$,
$\eta=0.02U$, $\mu=0.1U$ (a), and $\mu=0.9U$ (b).} \label{Fig_vi}
\end{figure*}
For the case of half-filling comparison with the data of Monte-Carlo
simulations \cite{Grober} shows that the spectra of the one-loop
approximation are closer to the results obtained at $T=0.33|t|$ than to
those derived for $T=0.1|t|$. \cite{Sherman06} The value $T=0.33|t|$ is
close to the superexchange constant $J$ for $t=-U/8$. For such
temperatures, the correlation length of the short-range
antiferromagnetic order is comparable to the intersite distance.
\cite{Shimahara} Thus, it can be concluded that the one-loop
approximation is more appropriate for short correlation lengths. This
conclusion is also corroborated by the shapes of the spectral function
in Fig.~\ref{Fig_iii}a which, as mentioned, are close to those at
half-filling. In this figure there are no indications that the unit
cell is doubled which manifests itself in a replication of some parts
of the quasiparticle dispersion with the period $(\pi,\pi)$. Such a
doubling is inherent in the antiferromagnetic order with a correlation
length which is much larger than the lattice spacing. This limitation
of the one-loop approximation is partly compensated by the fact, known
from experiment in cuprates \cite{Keimer} and from the $t$-$J$ model,
\cite{Sherman94} that the correlation length decreases rapidly with
departure from half-filling and becomes comparable to the lattice
spacing already at $1-\bar{n}\approx 0.1$ even for temperatures $T\ll
J$.

The spectral functions and quasiparticle dispersions calculated in the
one-loop approximation are close to those obtained in Monte-Carlo
simulations and cluster methods, provided that doping or temperature
ensure a short magnetic correlation length (cf.\ Fig.~\ref{Fig_iii}
with Figs.~9-11 in Ref.~\onlinecite{Grober} and Fig.~2 in
Ref.~\onlinecite{Kyung}). The most important differences are connected
with the fact that the one-loop approximation overestimates the
spectral intensity in the lower inner band near the momentum
$(\pi,\pi)$ and in the upper inner band near $(0,0)$. Since the Fermi
level crosses these bands with departure from half-filling, this leads
to underestimating (overestimating) of the electron concentration
\begin{equation}\label{concentration}
\bar{n}=\frac{2}{N}\sum_{\bf k}\int_{-\infty}^\infty{\rm
d}\omega\frac{A({\bf k}\omega)}{\exp(\beta\omega)+1}
\end{equation}
at hole (electron) doping. For example, for the parameters of
Figs.~\ref{Fig_iii}a and~b the concentrations calculated with the use
of Eq.~(\ref{concentration}) equal to 0.87 and 0.74, respectively.
However, from the comparison with the results of Monte-Carlo
calculations \cite{Grober} it can be concluded that the concentrations
have to be approximately 0.95 and 0.9, respectively. This is the reason
why the spectra in the above figures were labeled with the chemical
potential rather than with the concentration.
\begin{figure*}
\includegraphics[width=14cm]{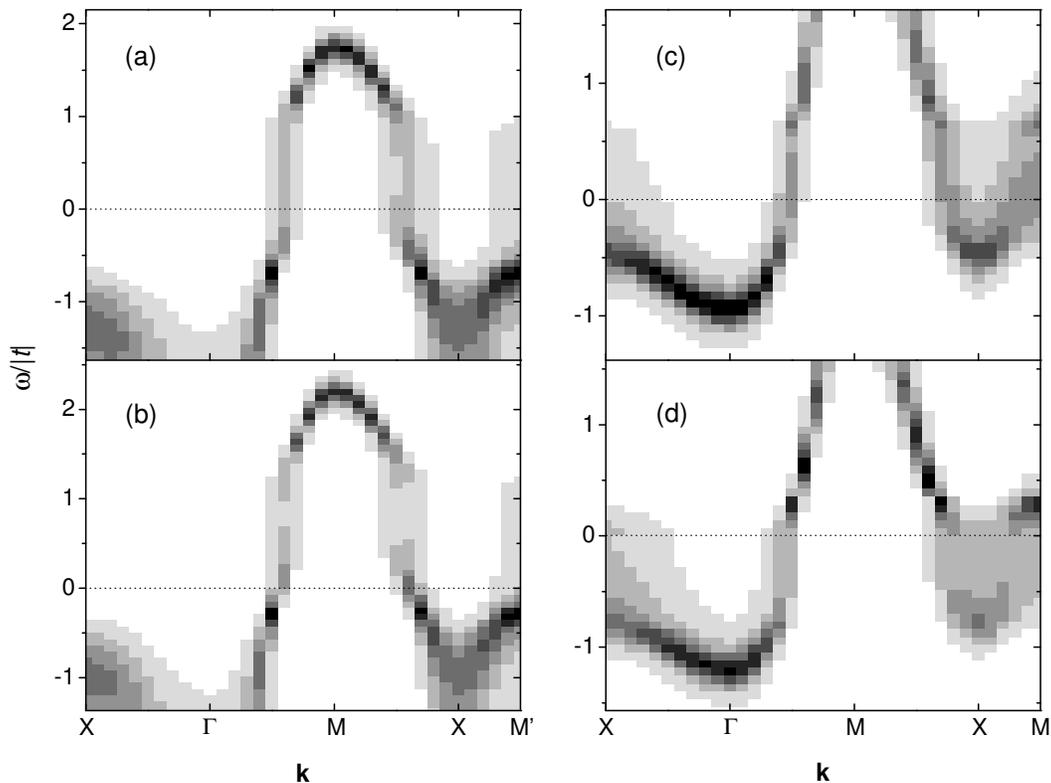}
\caption{The dispersion of maxima of the spectral function in the
$t$-$t'$-$U$ model on a 20$\times$20 lattice for $t=-U/8$, $t'/t=-0.3$,
$T=0.001U$, $\eta=0.02U$, $\mu=0.1U$ (a), $0.05U$ (b), $0.9U$ (c), and
$0.95U$ (d). Here darker areas correspond to larger intensities. The
points X, $\Gamma$, M, and M' correspond to the momenta $(\pi,0)$,
$(0,0)$, $(\pi,\pi)$, and $(\pi/2,\pi/2)$, respectively.}
\label{Fig_vii}
\end{figure*}

\section{The $t$-$t'$-$U$ model}
Now let us consider the $t$-$t'$-$U$ model with the initial electron
dispersion $t_{\bf k}=2t[\cos(k_x)+\cos(k_y)]+4t'\cos(k_x) \cos(k_y)$.
Both for the case of hole and electron doping the ratio $t'/t=-0.3$ of
the hopping constants for the next-nearest and nearest neighbors is
accepted. \cite{Korshunov} For both cases the spectral function and the
dispersion of quasiparticle peaks near the Fermi level are shown in
Figs.~\ref{Fig_vi} and~\ref{Fig_vii}. As in the case of the $t$-$U$
model, at half-filling the spectrum of the $t$-$t'$-$U$ model contains
four well separated bands. Also in analogy with the former model a
dispersive feature and a pseudogap appear near the Fermi level at a
certain level of doping. However, in the case of the $t$-$t'$-$U$ model
the obvious asymmetry of the hole and electron doping stands out -- the
new feature and the pseudogap are less pronounced in the case of
electron doping. This is also apparent from the comparison of
Figs.~\ref{Fig_vii}a and~c. Such behavior is a consequence of the
asymmetry in the filling dependence of ${\rm Im}K(\omega)$ -- for
identical offsets from $\mu=U/2$ the sharp minima which are similar to
those shown in Fig.~\ref{Fig_ii}b are more intensive in the case of
hole doping than for electron doping. This asymmetry is connected with
the contribution of the second term on the right-hand side of
Eq.~(\ref{simplified_ii}) to ${\rm Im}K(\omega)$.

Figure~\ref{Fig_vii} demonstrates the dispersions of the new band and
of the lower (upper) inner band in the case of hole (electron) doping,
the inner band being located above (below) the Fermi level. The
pseudogap between these bands becomes more pronounced with increasing
$\bar{n}$ in the case of electron doping (cf.\ Figs.~\ref{Fig_vii}c and
d). Similarly to the $t$-$U$ model, for the hole-doped case the
magnitude of the pseudogap observed in photoemission decreases with
increasing the hole doping (see Figs.~\ref{Fig_vii}a and~b), while for
the electron-doped case this magnitude increases with increasing the
electron doping (see Figs.~\ref{Fig_vii}c and~d).

Also as for the $t$-$U$ model, shapes of the spectral function and
quasiparticle dispersions calculated in the $t$-$t'$-$U$ model in the
one-loop approximation are similar to those obtained in the cluster
methods, provided that doping or temperature are high enough to ensure
a short magnetic correlation length (cf.\ Fig.~\ref{Fig_vi} with Fig.~2
in Ref.~\onlinecite{Senechal} and with Fig.~5a in
Ref.~\onlinecite{Kyung}). Again the most important difference between
these two groups of results is a larger spectral intensity in the upper
inner band near $(0,0)$ for the hole-doped case and in the lower inner
band near $(\pi,\pi)$ for the electron-doped case in the one-loop
approximation.

\section{Conclusion}
In the present work the diagram technique was used for the
investigation of the energy spectra of the $t$-$U$ and $t$-$t'$-$U$
Hubbard models at a departure from half-filling. The one-loop
approximation was applied which in the used diagram approach is a
successive improvement to the Hubbard-I approximation. In agreement
with results of Monte-Carlo simulations and cluster methods at
half-filling the obtained spectra of the models contain four bands. The
four-band structure persists also for some departure from half-filling.
Additionally in these conditions a new narrow band of quasiparticle
states arises near the Fermi level. The band energy is maximum near the
boundary of the magnetic Brillouin zone. The dispersion of the band is
much larger in the direction $(\pi/2,\pi/2)-(0,0)$ than along the
boundary of the magnetic Brillouin zone $(\pi,0)-(0,\pi)$. The width of
the band is of the order of the superexchange constant and decreases
with increasing doping. By these properties the new band resembles the
spin-polaron band of the $t$-$J$ model. For moderate doping the
intensities of maxima in the new band and in one of the inner bands of
the four-band structure decrease as the Fermi level is approached. As a
consequence a pseudogap arises in the spectrum near the Fermi level.
With hole doping the magnitude of the pseudogap observed in
photoemission decreases and eventually the pseudogap disappears in
agreement with experimental observations. With electron doping the
magnitude of the photoemission pseudogap increases. Shapes of the
spectral function and quasiparticle dispersions calculated in the
one-loop approximation are similar to those obtained by Monte-Carlo
simulations and cluster methods, provided that doping or temperature
are high enough to ensure a short magnetic correlation length.

\end{document}